\documentclass{article}
\usepackage{graphicx} 
\usepackage[utf8]{inputenc}
\usepackage{subfig}
\usepackage{indentfirst}
\usepackage{arydshln}

\usepackage{algpseudocode}
\usepackage{tablefootnote}

\usepackage{colortbl}
\usepackage{threeparttable}
\usepackage[normalem]{ulem}
\usepackage[colorinlistoftodos]{todonotes}
\usepackage{graphicx}
\usepackage{tabularx,colortbl}
\usepackage{makecell}
\usepackage{multicol}
\definecolor{LightBlue}{RGB}{140,186,252}
\newcolumntype{a}{>{\columncolor{LightRed}}c}
\usepackage{amsthm,amsmath,amssymb,mathtools}
\usepackage{algorithm}
\usepackage{enumerate}
\usepackage{enumitem}
\usepackage{bbm}
\usepackage{titlesec}
\usepackage{algcompatible}
\newcommand{\codecomment}[1]{\hfill \(\triangleright\) #1}
\usepackage{bigints}
\usepackage{setspace} 
\usepackage{verbatim} 
\usepackage{booktabs} 
\usepackage[width=0.9\textwidth]{caption}
\usepackage{fullpage}
\usepackage{hyperref}
\usepackage[toc,page]{appendix}
\usepackage{natbib}
\usepackage{booktabs}  
\usepackage{multirow}  
\usepackage{arydshln}  
\usepackage{array}     

\def\1{\mathbbm{1}}

\usepackage{tikz}
\usetikzlibrary{shapes,arrows}
\usetikzlibrary{calc}

\newtheorem{theorem}{Theorem}

\newtheorem{remark}{Remark}

\title{Adaptive Forward Stepwise Regression}
\author{Ivy Zhang\thanks{Dept. of Statistics, Stanford University; ivyzhang@stanford.edu} \ and Robert Tibshirani\thanks{Depts. of Statistics and Biomedical Data Science, Stanford University; tibs@stanford.edu} }
\date{\today}

\begin{document}

\maketitle

\begin{abstract}
This paper proposes a sparse regression method that continuously interpolates between Forward Stepwise selection (FS) and the LASSO.  When tuned appropriately, our solutions are much sparser than typical LASSO fits but, unlike FS fits, benefit from the stabilizing effect of shrinkage. Our method, \emph{Adaptive Forward Stepwise Regression} (AFS) addresses this need for sparser models with shrinkage. We show its connection with boosting via a soft-thresholding viewpoint and demonstrate the ease of adapting the method to classification tasks. In both simulations and real data, our method has lower mean squared error and fewer selected features across multiple settings compared to popular sparse modeling procedures. 
\end{abstract}

\section{Introduction}
Feature selection is essential for learning models that are both interpretable and predictive. Sparse regression can achieve this by identifying informative predictors and estimating their coefficients. These properties are especially attractive in the sciences where the goal is to understand what variables drive a response. As a result, sparse modeling remains an essential tool for modern practitioners.

We specifically consider the problem of simultaneous feature selection and prediction under linear models. Under the usual linear regression setting for a response, $y$, and random error $\epsilon$: 
\begin{equation}
y=X\beta+\epsilon,
\end{equation} with $\ y \in \mathbbm{R}^n, X \in \mathbbm{R}^{nxp}$. Without loss of generality, assume the data are centered so we do not include an intercept. We also assume the coefficient vector, $\beta$, is sparse, i.e., $\|\beta\|_0 = \sum_{k=1}^p \mathbbm{1}\{\beta_k \neq 0\}$ is small. Here, the $\ell_0$ pseudo-norm of $\beta$ being small tells us that only a small subset of the $p$ predictors are relevant. This assumption naturally motivates the Best Subset (BS) objective, 
\begin{equation}\label{eqn:bestsubset}
    \underset{\beta \in \mathbbm{R}^p}{\text{min }} \|y-X\beta\|^2+\lambda \|\beta\|_0,
\end{equation} which aims to optimally select a subset of features by introducing a $\ell_0$ penalty constraint to the ordinary least squares (OLS) objective. However, solving such a non-convex problem is NP-hard, leading to a preference for alternative methods \citep{NP}. 

One of the most popular methods for solving such a problem is the LASSO. This method minimizes the $\ell_1$ penalty objective
\begin{equation}\label{eqn:lasso}
     \underset{\beta \in \mathbbm{R}^p}{\text{min }} \|y-X\beta\|^2+\lambda \|\beta\|_1,
\end{equation} whose convex formulation makes it more computationally attractive \citep{lasso, BP}. Additionally, under certain regularity conditions, the LASSO is model selection consistent (ie, recovers the true support as $n\rightarrow \infty$) and has good prediction accuracy \citep{lassomodelconsistent, lasso_pred}. A closely related approach,  \textit{Least Angle Regression} (LAR), approximates the LASSO solution via a sequential procedure \citep{lars}. LAR begins by setting all coefficients $\hat{\beta}^{LAR}$ equal to 0 and initializing an empty active set, $\mathcal{A}=\emptyset$. At each step, LAR fits a model on the data subsetted for the selected variables from the previous step. It then selects the predictor most correlated with the residual. LAR then moves the coefficient for that predictor towards its OLS coefficient until another predictor achieves equal correlation with the current residual. Finally, it repeats the variable selection procedure and coefficient estimate in the subsequent steps until it computes the full coefficient path like the LASSO.

A drawback of the LASSO is its tendency to select excess noise variables after cross-validation of $\lambda$, leading to denser models. This behavior often arises because of the $\ell_1$ penalty's simultaneous role of shrinkage and model selection, especially when regularity conditions are not met \citep{sparsenet, twinboost, cvlasso}. A classical alternative to this is Forward Stepwise selection (FS). In one variant, FS starts with all coefficients set to zero and an empty active set, similar to LAR. It then uses the same selection criterion as well. Unlike LAR, FS produces coefficients that are the full OLS estimates fit on the selected features. While FS often yields sparser models when appropriately tuned, the absence of shrinkage results in higher variance coefficient estimates. Additionally, FS tends to perform less accurately than the LASSO in low signal-to-noise ratio (SNR) settings \citep{bestsubset}.

Recently, there has been ongoing development of penalty-based methods to achieve sparser models. For instance, the Relaxed LASSO (RLASSO) addresses some limitations of the LASSO by decoupling model selection from shrinkage \citep{rlasso}. This induces sparser estimates and improves prediction accuracy particularly in high SNR, while approximating the LASSO solution in low SNR scenarios. Other approaches explore non-convex optimization methods \citep{scad, MC}. For example, SparseNet uses a non-convex penalty that transitions between $\ell_1$ and $\ell_0$ regularization, bridging the LASSO and BS. In simulations, SparseNet yields similar or better predictive accuracy with fewer selected variables. However, unlike the LASSO, SparseNet does not readily generalize to non-Gaussian error models. See \cite{sparsenet} for details.

Others approached the problem through iterative methods like boosting,  synthetic data techniques, or independence learning. Boosting, as noted by \cite{reg}, acts as a regularization method for model estimation. \cite{lars} highlighted the connection between the LASSO and \cite{boosting}'s least squares boosting algorithm. \cite{l2boost} introduced $L_2$Boost, a variant aimed at consistently recovering the true regression function in high-dimensional and sparse settings. Meanwhile, \cite{stabl} proposed the Stabl algorithm, which applies a base regularization model, such as the LASSO, on both original and synthetic data samples. This method selects features that meet a ``reliability threshold" intended to reduce the number of false positive features selected. Another iterative method is the iterative sure independence screening (ISIS), which updates a set of important variables, selected based on marginal utility, conditional on the previous step \citep{sure}. 

This paper proposes Adaptive Forward Stepwise (AFS), a data-adaptive sparse regression method. We show that it bridges FS and LASSO, achieving the strengths of both when properly tuned. Empirically, we show that AFS's performance is more robust across a range of SNR, variable correlations, and dimensions, while also being more computationally efficient than several popular sparse modeling methods. In both simulations and real data, AFS also achieves higher sparsity than LASSO. Additionally, we demonstrate its application on real datasets and show that it can be easily adapted to classification tasks.

\subsection{Motivating example}

To illustrate the intuition behind our proposed method, we present three examples simulating a high-sparsity environment. The samples, $X_1,...,X_n$, and noise, $\epsilon_1,...,\epsilon_n$ are both independently drawn from Gaussian distributions as follows:
\begin{align*}
    X_1,..., X_n &\overset{\text{i.i.d}}{\sim} N(0, \Sigma), \ X_i\in \mathbbm{R}^p\ \\
    y_i&=X_i\beta + \epsilon_i, \ \epsilon_i \sim N(0, \sigma^2_\epsilon)
\end{align*}

Each covariate has the same variance and covariance, and only the first five covariates have non-zero coefficients. Specifically, $\forall k=1,...,p, k\neq j$, we set: 
\begin{align*}
    \Sigma_{k,k} &= \sigma^2_X, \ \Sigma_{j,k} = s_X \\
    \beta_1,..., \beta_5 &=2, \ \beta_6,...,\beta_p=0
\end{align*}

\begin{figure}[H]
   \centering
    \includegraphics[width=0.85\linewidth]{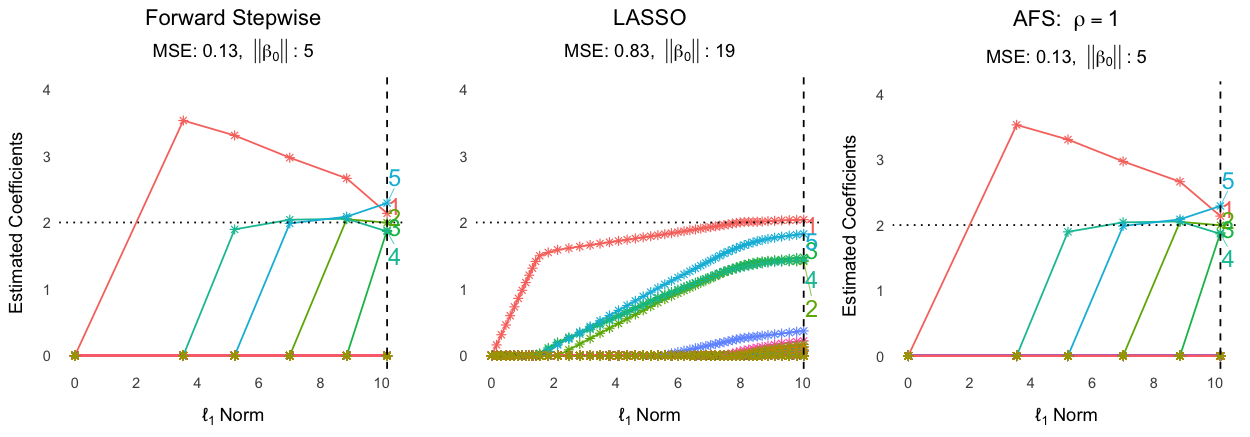}
    \caption{Figure \ref{fig:fig1}: SNR of 4.42, low correlation  of 0.06 between covariates, and $n\ll p$. The example illustrates the coefficient path of FS, LASSO, and AFS as a function of their $\ell_1$ norm. Paths of $\hat{\beta}_1,...,\hat{\beta}_5$ are annotated. Mean squared error, $\|X\hat{\beta}-\mu\|^2_2$, is notably lower for FS and AFS estimates compared to the LASSO. Both FS and AFS resulted in a sparser final model with $\|\beta_0\|=5$ than the LASSO, with $\|\beta_0\|=19$.}
    \label{fig:fig1}
\end{figure}

The first example has the settings (1) $n=100, \ p=120$, (2) high SNR of 4.42 and (3) low correlation of 0.06 between covariates. Figure \ref{fig:fig1} compares the coefficient paths of FS, LASSO, and AFS as a function of the $\ell_1$ norm. For FS and AFS, each knot represents one additional step taken in the respective algorithm. For the LASSO, each knot represents the next $\lambda$ in the $\lambda$-sequence evaluated. As each method progresses in their iterations/$\lambda$-sequence, the estimated model becomes more dense as seen by the entry of new coefficient path line segments. We use 10-fold cross-validation (CV) to select the hyperparameter for each method: number of steps for FS, $\rho$ and number of steps for AFS, and $\lambda$ for the LASSO. The vertical dashed line in Figure \ref{fig:fig1} indicates the CV solution along the path while the horizontal dotted line marks the true coefficient of the non-zero $\beta$'s.

In this high SNR, low correlation example, the CV solution identifies the true support for all three methods, but the LASSO introduces many more false positives with $\|\beta_0\|=19$, driving up the mean squared error (MSE) $\|X\hat{\beta}-\mu\|^2_2$. Furthermore, $\hat{\beta}^{LASSO}_1,...,\hat{\beta}^{LASSO}_5$ are further from the true $\beta$ values than FS. FS clearly performs better, with no false positives and all estimated coefficients close to $2$. The CV FS solution recovers the true support more accurately, with coefficient estimates much nearer to the true $\beta$ than the LASSO. Note that the CV solution for AFS matches that of FS. 

\begin{figure}[H]
      \centering
   \includegraphics[width=0.85\linewidth]{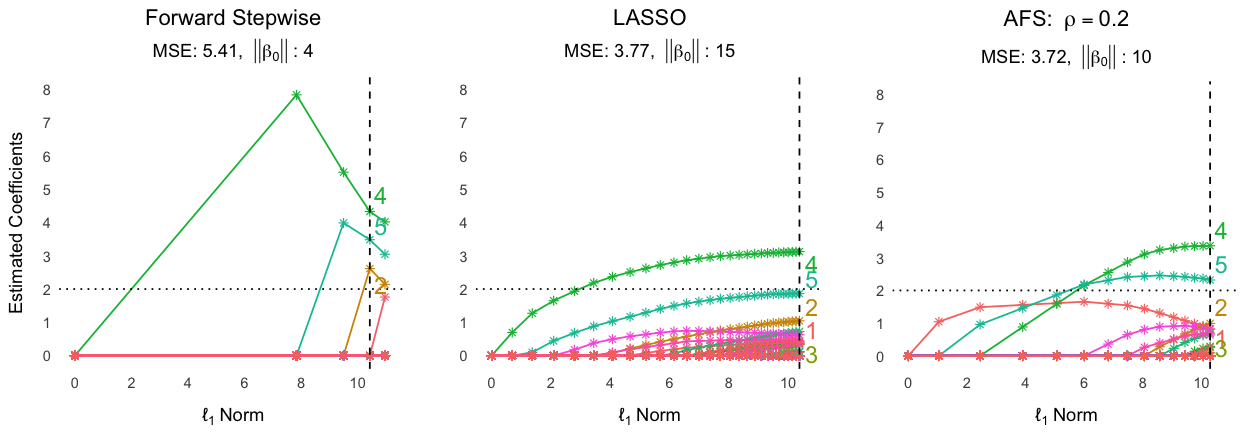}
   \caption{Figure \ref{fig:fig2}: SNR of 2.78, high correlation of 0.56 between covariates, and $n\gg p$ example illustrating the coefficient paths of FS, LASSO, and AFS as a function of their $\ell_1$ norm, with the layout as in Figure 1. The LASSO and AFS recover more of the true support than FS, while FS tends to overshoot the coefficient estimates. Both AFS and the LASSO benefit from shrinkage.}
   \label{fig:fig2}
\end{figure}

The second example illustrates the opposite end of the spectrum, with (1) $n=120, \ p=100$, (2) SNR of 2.78, and (3) high correlation of 0.56 between covariates where the LASSO outperforms FS. In this more challenging scenario with approximately half the SNR of Example 1 and high correlation, FS struggles to recover the full true support. As shown in Figure \ref{fig:fig2}, the LASSO recovers more of the true support and consequently achieves a significantly lower mean squared error, $\|X\hat{\beta}-\mu\|^2_2$, for its coefficient estimates. Unlike in the first example, the AFS CV solution resembles that of the LASSO more than FS. Here, AFS attains a similar MSE to the LASSO, with a much smaller model.

\begin{figure}[H]
    \centering
    \includegraphics[width=0.85\linewidth]{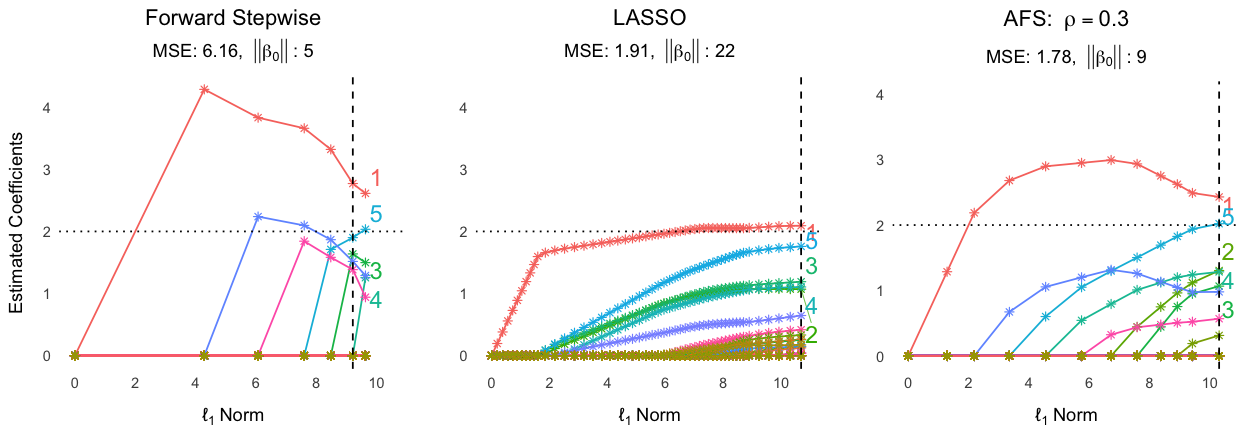}
    \caption{Figure \ref{fig:fig3}: SNR of 2.59, medium correlation  of 0.2 between covariates, and $n\ll p$. Unlike in Figure \ref{fig:fig1} where the AFS CV coefficient path matches that of FS or in Figure \ref{fig:fig2} where it is similar to that of the LASSO, this example shows an AFS coefficient path that represents a middle ground between FS and the LASSO.} 
    \label{fig:fig3}
\end{figure}

In the final example, we consider an intermediate regime with medium correlation. Here, we would ideally hope to achieve support recovery similar to the LASSO’s, but with fewer false positives. We would hope to also estimate non-zero coefficients closer to those in FS.  In this example, (1) $n=100, \ p=120$, (2) SNR of 2.59, and (3) medium correlation of 0.2. Figure \ref{fig:fig3} shows that AFS achieves this best of both worlds, resulting in the lowest MSE of the three with a model size less than half of the LASSO's. 

From the three examples, we see that AFS achieves robustness by adaptively constructing a model that resembles the clear winner between FS and the LASSO. Furthermore, it improves upon the LASSO by producing a sparser model while also improving upon FS by recovering more of the true support and applying shrinkage. 

\subsection{Outline of the paper}
In Section \ref{sec:afs}, we formally introduce AFS. Next, we illustrate the connection of our method to LAR, LASSO, and FS in Section \ref{sec:lars_fs_connection}. Additionally, we discuss the relationship between AFS and boosting, presenting AFS as a soft-thresholding procedure under orthogonal design in Section \ref{sec:boostingconnection}. Finally, we present the performance of our method compared to other popular sparse regression models across various simulations and real data examples in Sections \ref{sec:sim} and \ref{sec:data}. At the end of Section \ref{sec:data}, we outline an adaptation of the algorithm for any generalized linear model. We conclude with a summary of our contributions and a discussion in Section \ref{sec:discuss}.

\section{Adaptive Forward Stepwise Regression}
\subsection{The AFS algorithm}\label{sec:afs}

In this section, we introduce our method, Adaptive Forward Stepwise, to address the need for sparser models. Consider the linear regression setting for a response, $y$: 
\begin{equation}\label{afs_step0}
y=X\beta+\epsilon, \ \epsilon \sim N(0, \sigma^2I),
\end{equation} with $\ y \in \mathbbm{R}^n, X \in \mathbbm{R}^{nxp}$. Without loss of generality, assume the data are centered so we do not include an intercept. Suppose $\beta$ is sparse, i.e., the $\ell_0$ pseudo-norm of $\beta$, $\|\beta\|_0 = \sum_{k=1}^p \mathbbm{1}\{\beta_k \neq 0\}$ is small. Since this tells us only a small number of variables are relevant, our goal is to estimate a sparse model. 

Our proposed iterative method starts at step $m=0$ with setting all estimated coefficients, 
$$\hat{\beta}_{0,\rho}^{AFS}=0,$$ and an empty active set, $|\mathcal{A}|=0$, representing no selected variables. We also fix our shrinkage parameter, $\rho \in (0,1]$, and maximum number of steps, $M$, to iterate on. 

At each step, $m$, we choose the $j$th variable which maximizes the inner product between the vector, $x_j$, and the residual from the estimated coefficients. 
\begin{equation}\label{afs_criteria}
j^*_m= \underset{j \in \{1,..,p\}}{\text{argmax }} |x_j^\intercal (y- X\hat{\beta}^{AFS}_{m-1,\rho} )|.
\end{equation}

In the first step, $j^*_m= \text{argmax}_j |x_j^\intercal y|$ since $\hat{\beta}_{0,\rho}^{AFS}=0$ . For subsequent steps, we can interpret $j^*_m$ as the variable that maximizes the correlation with the response, after projecting out the contribution from the selected variables. Equivalently, we can think of $j^*_m$ as the variable that minimizes the angle between the $x_j$ and the residual, an interpretation which will help us make the connection with \textit{Least Angle Regression} in Section \ref{sec:lars_fs_connection}.

Once the selected variable is added to our active set, we estimate the OLS coefficients using the selected subset of the data:
\begin{equation}\label{nu_step}
    \hat{\nu}_m = (X_{\mathcal{A}_{m-1}}X_{\mathcal{A}_{m-1}})^{-1}X_{\mathcal{A}_{m-1}}^\intercal y.
\end{equation} In the case when $n\ll p$, we can add an early stopping rule, which ends the algorithm when the $\ell_1$ norm of the AFS coefficients exceeds the largest of the LASSO coefficient $\ell_1$ norm, i.e., $\|\hat{\beta}^{AFS}\|_1 > \underset{\lambda} {\text{max }}\|\hat{\beta}^{LASSO}(\lambda)\|_1$. This ensures that Eqn. \ref{nu_step} is well defined.

Finally, we define the AFS coefficient as 
\begin{equation}
    \hat{\beta}^{AFS}_{m,\rho}=(1-\rho) \hat{\beta}^{AFS}_{m-1,\rho} + \rho \hat{\nu}_m,
\end{equation} which applies a shinkage to both the previous step's estimated coefficients and the current step's estimated OLS coefficient. We summarize our method in Algorithm \ref{alg:afs}.

Algorithm \ref{alg:afs} produces a path of coefficients across $m$ which can be estimated using various resampling or covariance penalty methods to minimize the true test error. However, since commonly used covariance penalty methods like Mallow's $\text{C}_\text{p}$, AIC, and BIC depend on the degrees of freedom (df) of the estimator, we recommend selecting $\rho$ via cross-validation. This is for two main reasons: (1) accuracy and (2) computational efficiency. The estimator's dependence on $y$ for covariate selection makes it a non-linear function of $y$, complicating the estimation of df. While bootstrap methods for estimating df are often computationally prohibitive with large datasets, the traditional df approximation as the trace of the hat matrix $M$, given by $My=X\hat{\beta}$, can be highly crude for large $\rho$. Empirical results highlighting these issues are detailed in Appendix \ref{app:dof}.

\begin{algorithm}[H]
\caption{Adaptive Forward Stepwise}\label{alg:afs}
\begin{algorithmic}[1]
    \State Initialize all $p$ AFS coefficients $\hat{\beta}^{AFS}_{0,\rho} = 0$ and active set, $\mathcal{A}= \{\emptyset\}$. 
    \State For the following parameters, set
        \State \quad   $M$, the number of iterations, large \codecomment{Choose by CV}
        \State \quad  $\rho\in (0,1]$, the stepsize \codecomment{Choose by CV}
        \State \quad  $h= \underset{\lambda} {\text{max }}\|\hat{\beta}^{\text{LASSO}}(\lambda)\|_1$ , the maximum allowable $\ell_1$ norm
     \State While $m < M$ and $\|\hat{\beta}^{AFS}_{m,\rho}\|_1 < h$, let
        \State \quad  $ j^*_m= \underset{j \in \{1,..,p\}}{\text{argmax }} |x_j^\intercal (y- (\hat{\beta}^{AFS}_{m-1,\rho})^\intercal  X )|$ \codecomment{Select most correlated variable with current residuals} 
        \State \quad  $A_m=A_{m-1} \cup j^*_m$ \codecomment{Update active set}
        \State \quad  $\hat{\nu}_m= \hat{\beta}^{OLS}_{\mathcal{A}_m}$, the OLS coefficients of $y$ on the active set $\mathcal{A}_m$ \codecomment{Compute OLS coefficients}
        \State \quad $\hat{\beta}^{AFS}_{m,\rho}=(1-\rho) \hat{\beta}^{AFS}_{m-1,\rho} + \rho \hat{\nu}_m$ \codecomment{Update AFS coefficients} 
\end{algorithmic}
\end{algorithm}

\begin{remark}
    The structure of the algorithm allows us to easily conduct post-selection inference per the framework of \cite{lee_postselection}. Formally, we wish to conduct the following test: \begin{align*} H_0: v_k^\intercal  \theta = 0 \end{align*} conditional on the chosen $\mathcal{A}_m$ at step m. I.e., we are testing the coefficient of the variable, k, selected at step m is 0. In the case when $p < n$, their framework allows us to perform inference on the \textit{true population} $\beta_k$ by defining $v_k = (X^\intercal X)^{-1}X^\intercal e_k$, where $e_k$ is the $k$th standard basis vector. In the case when $p > n$, note that we can only perform inference on the submodel $\beta$ i.e., $v = (X_{A_{k}}^\intercal X_{A_{k}})^{-1}X_{A_{k}}^\intercal e_k $. 
Details can be found in Appendix \ref{app:inf}.
\end{remark}

\subsection{Connection to LAR, the LASSO, and Forward Stepwise}\label{sec:lars_fs_connection}

\begin{figure}[H]
    \centering
    \includegraphics[width=0.65\linewidth]{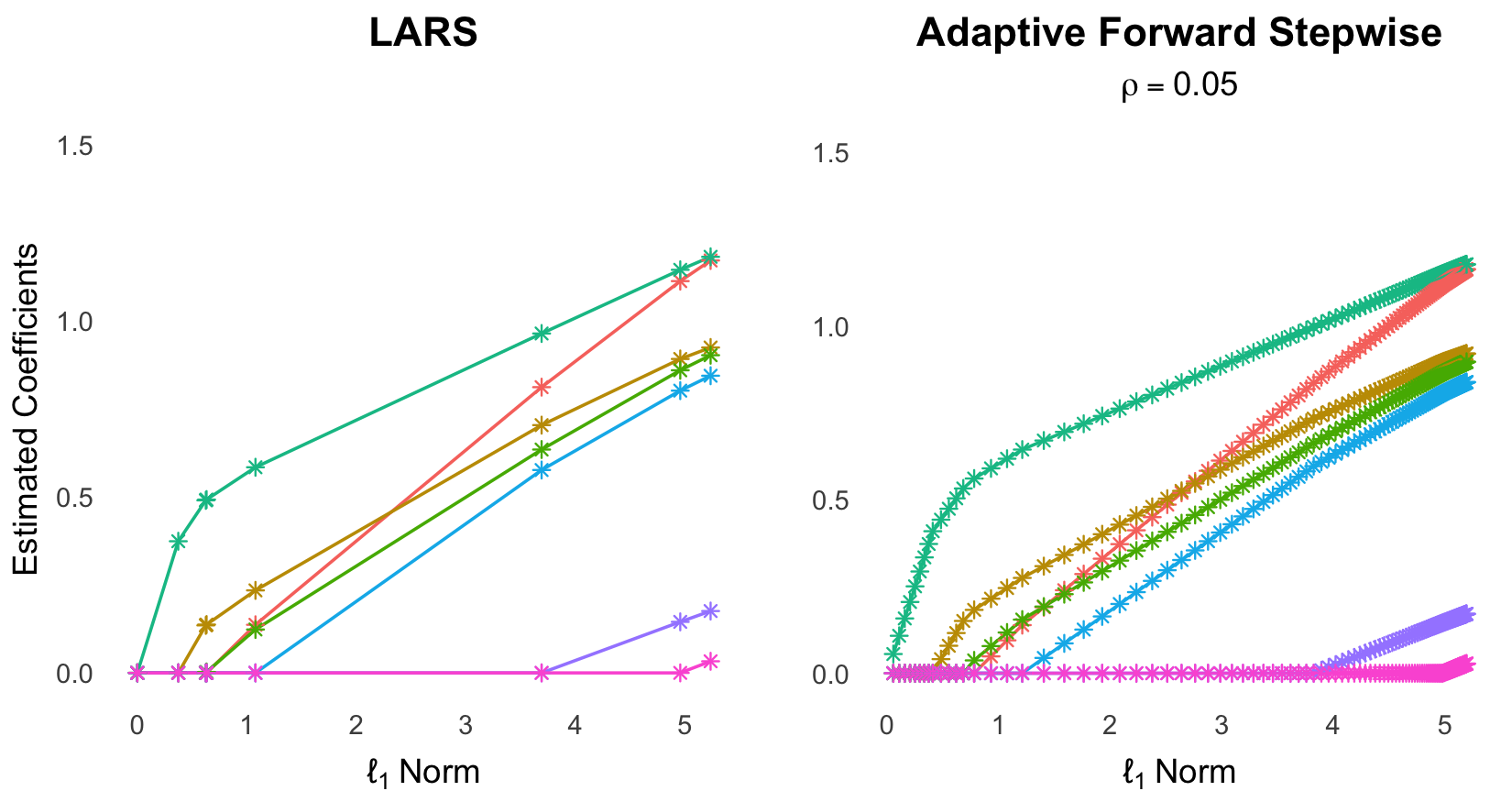}
    \caption{Figure \ref{fig:lars_afs}: Coefficient paths for LAR and AFS ($\rho=0.05$) from simulation of $n=100, p=7$. AFS takes many small steps to approximately trace out each line segment that make up the solution path of LAR. The AFS plot is truncated to exclude solutions with $\|\hat{\beta}^{AFS}\|_1 > 5.25$, the $\ell_1$ norm of the last step of the LAR coefficient path.}
    \label{fig:lars_afs}
\end{figure}

We now formally show that AFS interpolates between FS and LAR. For $\rho=1$, we have the FS procedure, which sequentially adds $j_m^*$ and fits $\hat{\beta}^{FS}_m=\hat{\beta}^{OLS}_{\mathcal{A}_m}$. For $\rho\rightarrow 0$, interestingly, we recover the LAR solution path as seen in Figure \ref{fig:lars_afs}. The plot shows the solution path of LAR and AFS for a simulation with $X_1,..,X_7$ drawn $i.i.d$ from a standard Gaussian distribution and $y = X\beta+N(0,1)$, with $\beta_1,.., \beta_5 = 1, \ \beta_6, \beta_7 = 0$. As we can see, for this small choice of $\rho=0.05$, AFS takes many small steps and traces out a similar path to LAR.

This result, formalized in Theorem \ref{theorem:thm1}, relies on the the fact that both methods use the same criterion for selecting a variable to add to the active set and the piecewise linear nature of their solution paths. The details of the proof can be found in Appendix \ref{app:thm1}.
\begin{theorem}\label{theorem:thm1}
Let $t=\|\hat{\beta}\|_1$ and assume there are no ties in variable selection for both procedures. Then
\begin{align*}
    \underset{\rho \rightarrow 0}{\text{lim }} \hat{\beta}^{\text{AFS}}_{\rho}(t) = \hat{\beta}^{\text{LAR}}(t) 
\end{align*}
\end{theorem}

This result suggests the following modification to Algorithm \ref{alg:afs} to approximately recover the LAR coefficients in practice: iteratively reduce $\rho$ in each step until the selection criterion selects a new variable. Since LAR approximates the LASSO when there is no change in any non-zero coefficient sign, this also approximately recovers the LASSO. In the case of a sign change, we can add one further modification, as detailed in Appendix \ref{lasso_mod}.

\subsection{Connection to Boosting and a Soft Thresholding Viewpoint}\label{sec:boostingconnection}

In this section, we provide another perspective on AFS to better understand its behavior. The fitting of a model based on residuals from the previous iteration suggests a possible connection with boosting. This connection is made clear from a soft-thresholding viewpoint. For the remainder of this section, we consider an orthogonal design, i.e. $X^\intercal X=I_p$, to gain some more intuition on our method.

At each iteration, $m$, the $j$th AFS coefficient is given by

\begin{equation}\label{eqn: ortho_betaj}
    \hat{\beta}^{AFS}_{j, \rho}(m) =\sum_{i=0}^{m-1} (x_j^\intercal y\rho)(1-\rho)^{i-k_j+1}= \hat{\beta}^{OLS}_{j}(1-(1-\rho)^{m-k_j+1}),
\end{equation} where $k_j$ is the iteration at which $x_j$ enters the active set and $\hat{\beta}^{OLS}_j$ is the OLS coefficient for the $j$th variable fit on the full design matrix. The first equality follows from the definition of $\hat{\beta}^{AFS}_{j, \rho}$ since the columns of $X$ are orthogonal. The second equality follows from evaluating the geometric sequence. The exponent, $\ell_{j,m} \coloneqq m-k_j+1$, denotes the number of times the $j$th variable has been in the active set by step m, inclusive. Therefore, the AFS coefficient for each variable is the OLS coefficient shrunk by a factor $1-(1-\rho)^\ell_{j,m}$, which decreases in $m$. Using this representation of the AFS coefficients helps us arrive at the approximate soft-thresholding estimator in Theorem \ref{theorem:thm2}.  

\begin{theorem}\label{theorem:thm2}
    Define the soft-thresholding estimator 
    \begin{equation*}\label{eqn: soft}
      \hat{\beta}^{ST}_{j}(\lambda) =
    \begin{cases}
      \hat{\beta}^{OLS}_j - \lambda & \text{if } \hat{\beta}^{OLS}_j \geq \lambda\\
      0 & \text{if } |\hat{\beta}^{OLS}_j| < \lambda\\
      \hat{\beta}^{OLS}_j + \lambda & \text{if } \hat{\beta}^{OLS}_j \leq -\lambda
    \end{cases}       
 \end{equation*} There exists some threshold $\lambda_{j,m}(\rho) \in [c(1-\rho)^{2\ell_{j,m}}, c(1-\rho)^2)$ for $c=\sqrt{1-(1-\rho)^2}$ such that Equation \ref{eqn: ortho_betaj} is well approximated by a soft-thresholding estimator with threshold $\lambda_{j,m}(\rho)$ for small $\rho$: 
    \begin{equation*}
      \underset{\rho \rightarrow 0}{\text{lim }} | \hat{\beta}^{AFS}_{j,\rho}(m) - \hat{\beta}^{ST}_{j}(\lambda_{j,m}(\rho))| = 0
\end{equation*}
under an orthogonal $X$. 
\end{theorem} 

\begin{figure}[H]
    \centering
    \includegraphics[width=0.85\linewidth]{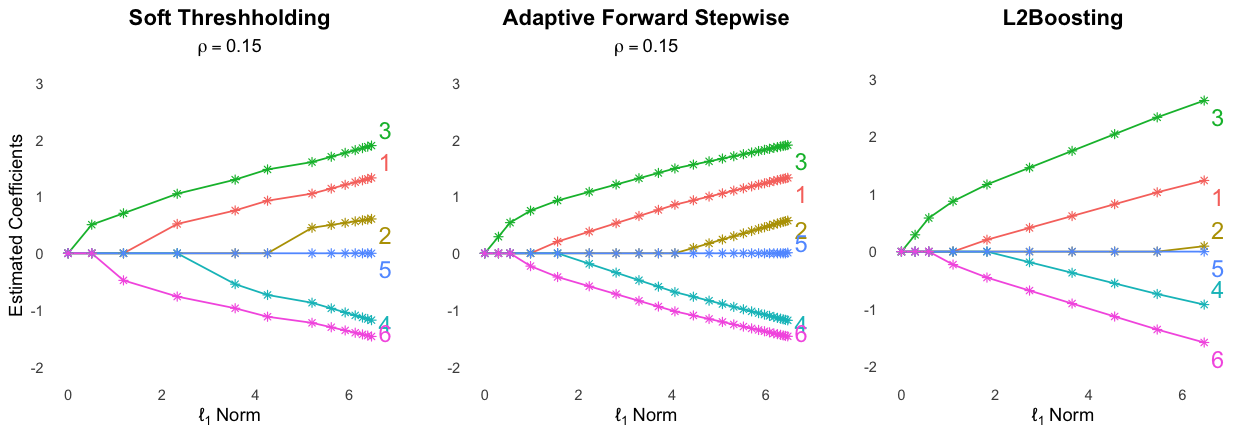}
    \caption{Figure \ref{fig: boost}: Coefficient paths for the approximate soft thresholding estimator \ref{eqn: soft} and AFS at $\rho=0.15$ compared to L2Boosting. Simulation for $n=100, p=6$ under an orthogonal design matrix $X$. }
    \label{fig: boost}
\end{figure}

We now compare this to boosting with a linear estimator under squared error loss, L2Boosting. Firstly, the L2Boosting analog of the right hand side of Equation \ref{eqn: ortho_betaj} is $$\hat{\beta}^{Boost}_{j,m} = \hat{\beta}_{j}^{OLS}(1-(1-\nu)^{\tilde{k}_j})$$ for step size $\nu$ and $\tilde{k}_j$ number of times variable $j$ has been chosen by iteration m. Since $\ell_{j,m} \geq \tilde{k}_j$, AFS will produce more shrunken coefficients when the same step sizes are chosen. Second, from the details of Theorem \ref{theorem:thm2} (see Appendix \ref{app: thm2}), we see that the residual sum of squares (RSS) for AFS is decreasing in $m$ and the difference in RSS decreases in $m$ by a factor of $(1-\rho)^2$ just like L2Boosting. This further connects the behavior of the AFS and boosting algorithms, illustrated in Figure \ref{fig: boost}.

\section{Simulation studies}\label{sec:sim}

In this section, we present simulations across various SNR, correlation between covariates, and dimension settings to compare the results of AFS, FS,  LASSO, RLASSO, SparseNet, and Stabl\footnote{\texttt{bestsubset::fs()} and \texttt{sparsenet::sparsenet()} in R only supports the Gaussian model so we do not compare their performance in the binary classification task of Figure \ref{fig: binomial_res}}. For each simulation, we draw data from a Gaussian model, apply CV to select any hyperparameters, and refit the model on the full dataset. No test-training split was used since the true mean, $\mu = X\beta$, is known. Each simulation sets $\beta_1,...,\beta_5 = 2$ and $0$ otherwise.

\subsection{Performance}

We now provide detailed results to compare the performance of AFS and other sparse regression methods. We consider regimes covering combinations from the following settings: SNR of $0.5$ (low), $1.0$ (medium), $1.5$, (medium), and $2.0$ (high); correlation of $0$, $0.15$ (low), and $0.6$ (high); $n=100$ and $n=120$; $p=100$ and $p=120$. 

The low, medium, and high SNR levels used here are based on the analysis in \cite{bestsubset} to reflect SNRs most commonly found in real data. The figures below show the median MSE, $\|X\hat{\beta}-\mu\|^2_2$ and FPR, proportion of false positive features selected, for each configuration. Error bars represent $\pm 1$ standard deviation across 50 trials. In the $n=120, p=100$ setting, AFS achieves one of the lowest median MSEs, while no other method consistently demonstrates top performance across all SNR levels. This demonstrates that AFS is a more robust compared to other popular methods, while producing highly sparse models. However, we note that in the high dimensional, high correlation setting, RLASSO outperforms AFS in robustness and sparsity.

\begin{figure}[H] 
    \captionsetup{width=\linewidth}
    \centering
    \includegraphics[width=0.8\linewidth]{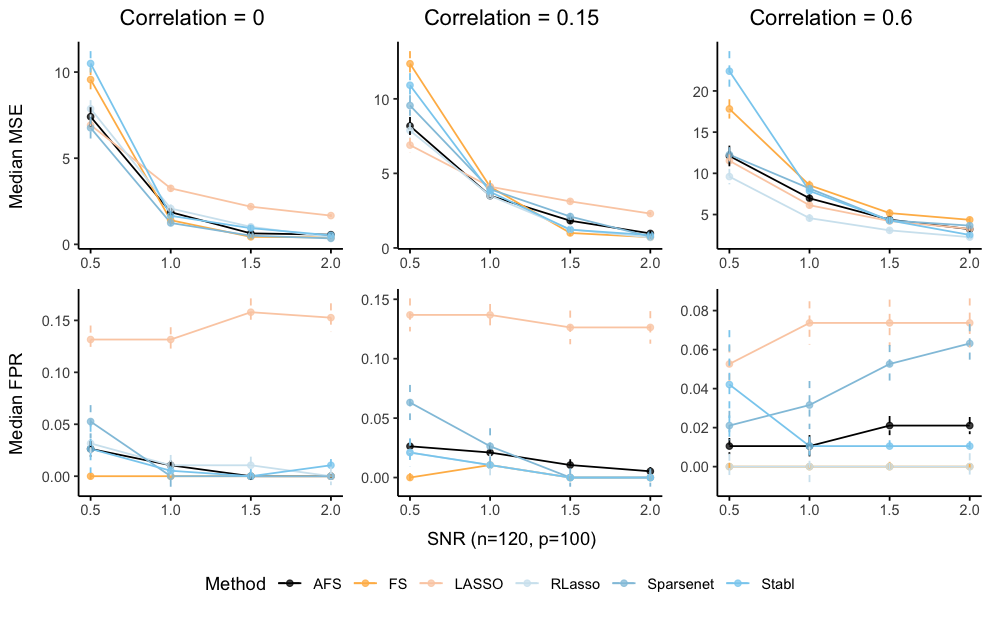}
    \caption{Figure \ref{fig: n120}: $n>p$ simulation results across $50$ trials. Error bars represent $\pm 1$ standard deviation. Unlike other methods, AFS performance is robust across various configurations while maintaining high sparsity.}
    \label{fig: n120}
\end{figure}

\begin{figure}[H] 
    \centering
    \captionsetup{width=\linewidth}
    \includegraphics[width=0.8\linewidth]{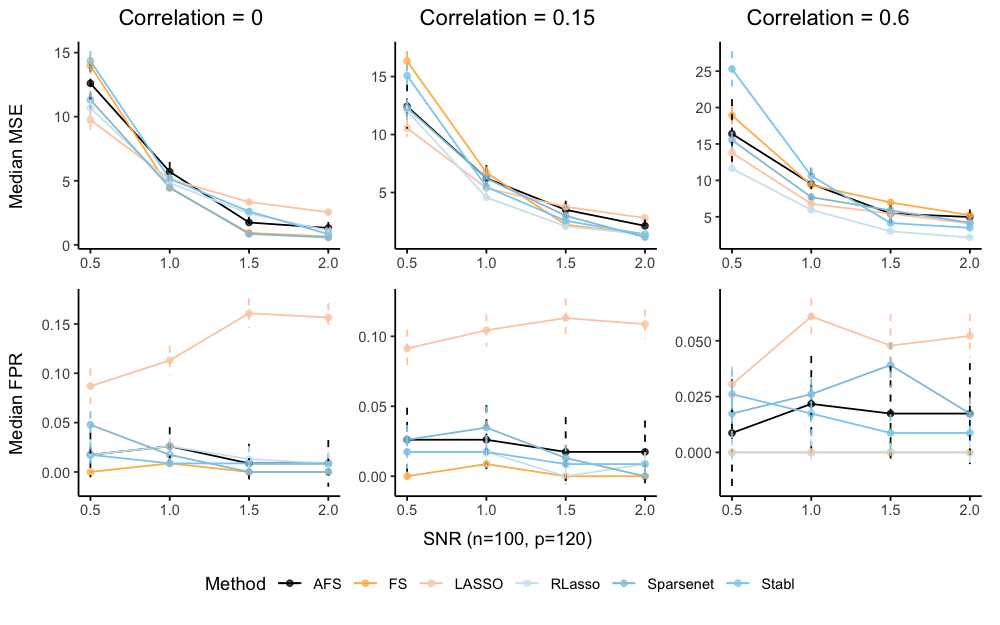}
    \caption{Figure \ref{fig: n100}: $n<p$ simulation results across $50$ trials. As in the $n>p$ setting, AFS MSE does not fluctuate as strongly as other methods across SNR for a fixed data correlation--for example, under $0$ and $0.15$ correlation, the LASSO has the lowest MSE under $0.5$ SNR but highest MSE under $1.5-2$ SNR. However, compared to the low dimension case, AFS struggles more here.}
    \label{fig: n100}
\end{figure}

\subsection{Computation time}
\begin{figure}[H]
    \centering
    \includegraphics[width=0.65\linewidth]{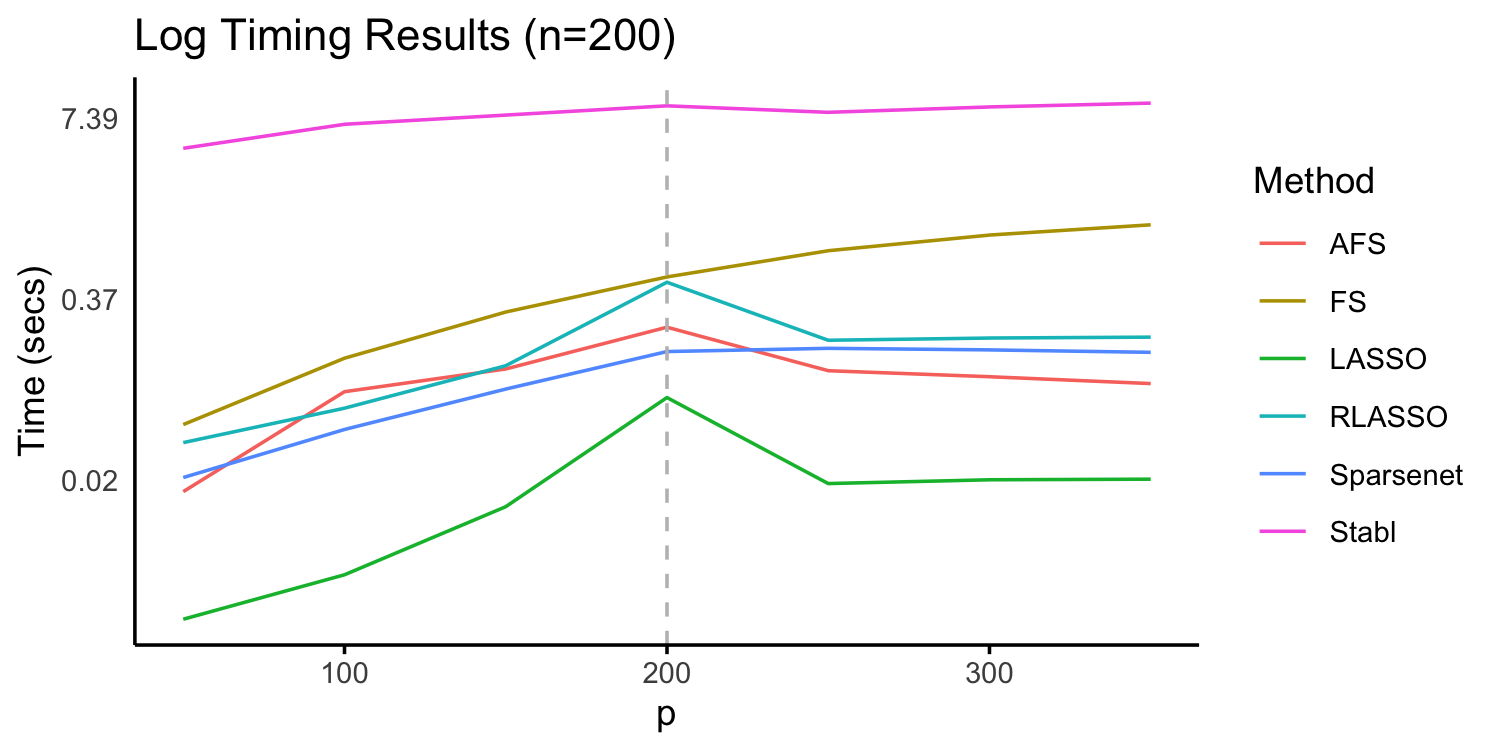}
    \caption{Fig \ref{fig:timing1}: Time in seconds to fit each model under $n=200$ and varying $p$, averaged across $50$ trials. AFS (computations in \texttt{R}), demonstrates computational efficiency similar to that of SparseNet (computations in \texttt{fortran}), beating Stabl (computations in \texttt{R}), FS (computations in \texttt{C}), and RLASSO (computations in \texttt{C}). The LASSO (computations in \texttt{C}) remains the most computationally efficient across all $p$ }
    \label{fig:timing1}
\end{figure}

To assess computational time, we measured the runtime required for each method to fit the full coefficient path (excluding Stabl, which lacks a coefficient path) for $n=200$ and varying $p$. Figure \ref{fig:timing1} presents the average computation time over $50$ trials with the following settings: (1) $0.15$ correlation (2) $1.0$ SNR and (3) $\beta_1,...,\beta_5=2$ and $0$ otherwise. To allow for better comparison, hyperparameter selection via CV was excluded from the timing. However, we note that computations for LASSO/RLASSO (\texttt{glmnet}) and FS (\texttt{bestsubset}) are based in \texttt{C}, SparseNet in \texttt{fortran}, and AFS and Stabl in \texttt{R}. AFS shows lower computational expense than most methods due to efficient matrix inverse updates at each iteration. Stabl, the least efficient method, incurs added costs from incorporating synthetic data, while the LASSO remains faster than all methods across all dataset sizes. 

\section{Real data examples}\label{sec:data}

\begin{figure}[H]
    \centering
    \captionsetup{width=\linewidth}
    \includegraphics[width=\linewidth]{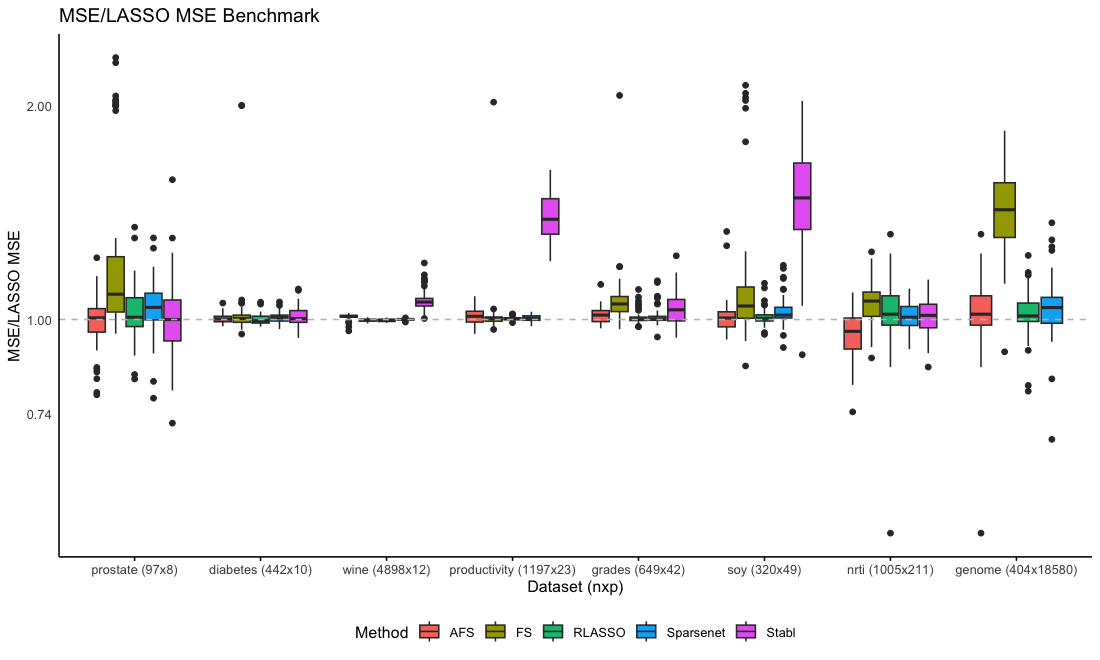}
    \caption{Figure \ref{fig: data}: MSE relative to LASSO MSE results from 50 trials on a log scale. Stabl was unable to run for the genome dataset due to insufficient memory allocation for vector storage in \texttt{R}. }
    \label{fig: data}
\end{figure}

\begin{figure}[H]
    \centering
    \captionsetup{width=\linewidth}
    \includegraphics[width=\linewidth]{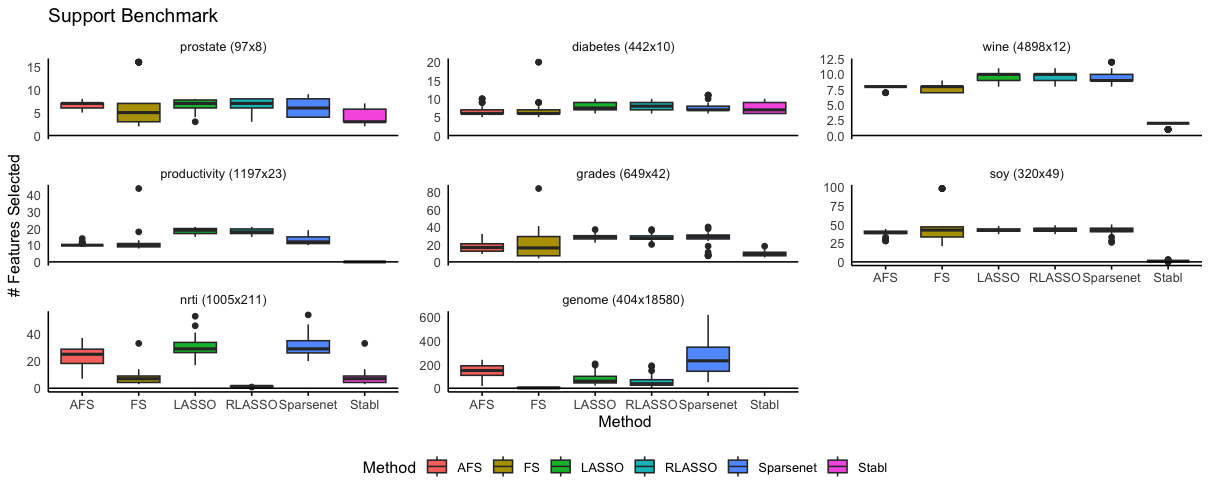}
    \caption{Figure \ref{fig:data_support}: Number of features selected in the final model. AFS generally selects fewer features than other methods on smaller datasets, while maintaining competitive performance, except in the high dimensional, high correlation genome dataset. There, RLASSO has the best combination of sparsity and MSE.}
    \label{fig:data_support}
\end{figure}

In this section, we compare the performance of AFS and other sparse regression methods across eight publically available datasets: prostate ($n=97, p=8$) \citep{eslr}, diabetes ($n=442, p=10$) \citep{lars}, wine ($n=4898, p=12$) \citep{wine}, productivity ($n=1197, p=23$) \citep{productivity}, student grades ($n=649, p=42$) \citep{grades}, soy ($n=320, p=49$) \citep{soy}, energy ($n=19735, p=28$) \citep{energy}, nrti ($n=1005, p=211$) \citep{nrti}, and genome ($n=404$, $p=18580$) \citep{genome}. In the first plot, Figure \ref{fig: data},  we present MSE on a held-out test set over $50$ trials of a 15-85\% test-train split, relative to that of the LASSO, on a log scale. Hyperparameters for AFS, FS, SparseNet, RLASSO, and the LASSO are selected by $10$-fold CV. The second plot, Figure \ref{fig:data_support}, displays the number of features selected in the final models.

From these real data examples, we observe that AFS achieves comparable to or better performance than the LASSO and other methods on datasets with fewer than $50$ covariates, while yielding sparser models. Some methods, such as FS, achieve higher sparsity than AFS but compromise predictive performance. AFS particularly excels on the larger nrti dataset. However, it faces more challenge in the high-dimensional, high-correlation setting of the genome dataset, aligning with simulation results in Section \ref{sec:sim}.

\subsection{Extension to Generalized Linear Models}

One advantage of AFS is the adaptability of the method to any generalized linear model (GLM) by modifying the selection criterion with the appropriate score function and fitting $\hat{\beta}^{GLM}$ instead of $\hat{\beta}^{OLS}$. For example, a binary classification task using a logistic regression can be adapted into the AFS framework as in Algorithm \ref{alg:afs_logit}. Compared to Algorithm \ref{alg:afs}, only lines $7-9$ differ.

\begin{algorithm}[H]
\caption{Adaptive Forward Stepwise - Logit Link Modifications}\label{alg:afs_logit}
\begin{algorithmic}[1]
    \State Initialize all $p$ AFS coefficients $\hat{\beta}^{AFS}_{0,\rho} = 0$ and active set, $\mathcal{A}= \{\emptyset\}$. 
    \State For the following parameters, set
        \State \quad   $M$, the number of iterations, large \codecomment{Choose by CV}
        \State \quad  $\rho\in (0,1]$, the stepsize \codecomment{Choose by CV}
        \State \quad  $h= \underset{\lambda} {\text{max }}\|\hat{\beta}^{\text{LASSO}}(\lambda)\|_1$ , the maximum allowable $\ell_1$ norm
     \State  While $m < M$ and $\|\hat{\beta}^{AFS}_{m,\rho}\|_1 < h$, let
        \State \quad $ j^*_m= \underset{j \in \{1,..,p\}}{\text{argmax }} \bigg|x_j^\intercal \left({y}- \frac{\text{exp}((\hat{\beta}^{AFS}_{m-1, \rho})^\intercal  X)}{(1+\text{exp}((\hat{\beta}^{AFS}_{m-1, \rho})^\intercal  X )}\right)\bigg|$ \codecomment{Select most correlated variable with current residuals} 
        \State \quad $A_m=A_{m-1} \cup j^*_m$ \codecomment{Update active set}
        \State \quad $\hat{\nu}_m= \hat{\beta}^{Logistic}_{\mathcal{A}_m}$, the logistic regression coefficients of y on the active set, using $\hat{\beta}^{Logistic}_{\mathcal{A}_{m-1, \rho}}$ as warm start
        \State \quad $\hat{\beta}^{AFS}_{m,\rho}=(1-\rho) \hat{\beta}^{AFS}_{m-1, \rho} + \rho \hat{\nu}_m$ \codecomment{Update AFS coefficients} 
\end{algorithmic}
\end{algorithm}

\begin{figure}[H]
    \centering
    \includegraphics[width=0.75\linewidth]{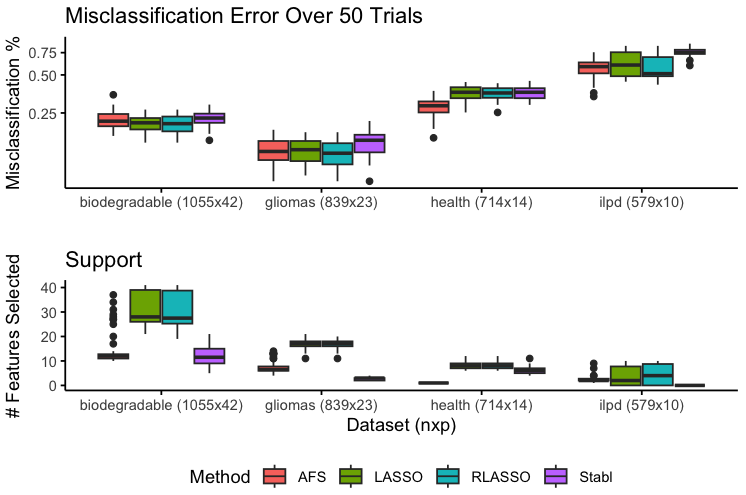}
    \caption{Fig \ref{fig: binomial_res}: Comparison of AFS,  LASSO, RLASSO, and Stabl for a binary classification task across UCI datasets. Test misclassification \% on the log scale and number of non-zero coefficients (support) of the final model are reported.}
    \label{fig: binomial_res}
\end{figure}

In Figure \ref{fig: binomial_res}, we present an application of AFS for binary classification using four datasets from the UCI repository \citep{uci_biodegrade, uci_glioma, uci_health, uci_ilpd}. For each dataset, we compare the performance of AFS,  LASSO, RLASSO, and Stabl over 50 trials. A 15-85\% train-test split was applied, and 10-fold CV was used to select hyperparameters for AFS, LASSO, and RLASSO on the training set. The misclassification percentage (log-scaled) on the test set and the final selected model size are reported.

Our experiments show that AFS generates significantly sparser models than the LASSO and RLASSO, while achieving comparable or improved accuracy. Although Stabl sometimes produced even sparser models than AFS, this often came at the cost of reduced accuracy. Notably, on the most challenging dataset—where all methods performed worse than random chance—AFS still outperformed the other approaches and showed lower variance across the 50 trials.

\section{Discussion}\label{sec:discuss}

We propose a method, Adaptive Forward Stepwise, to address the need for sparser solutions than the LASSO while balancing predictive performance and computational efficiency. AFS is a sparse regression method that bridges Forward Stepwise and the LASSO. Our method produces sparser solutions than the LASSO under appropriate tuning while still allowing for shrinkage, unlike FS. Across numerous simulations under varying signal-to-noise ratios and correlation structures, AFS produces robust performance. Comparatively, other methods experienced large variation of MSE and sparsity across different settings. Similarly, in numerous real data examples, AFS matches or outperforms the LASSO, FS, SparseNet, RLASSO, and Stabl. Our method is also easily modifiable to classification tasks and can be implemented in less time than several other methods. While AFS excels in maintaining one of the lowest MSEs across different settings, it encounters challenges in high-correlation, $n \ll p$ scenarios. An implementation of our method as a Python and R package is forthcoming.

\section*{Acknowledgments}
R.T. was supported by the NIH (5R01EB001988- 16) and the NSF (19DMS1208164). We thank Asher Spector, James Yang, and Tim Morrison for insightful discussions and draft feedback.

\bibliography{ref.bib}

\newpage

\section*{Appendix}
\appendix
\section{Proof of results}
\subsection{Proof of Theorem \ref{theorem:thm1}}\label{app:thm1}
\begin{proof}
We first show that for any fixed $\epsilon > 0$, there exists a $\rho \rightarrow 0$ such that for any $t$, 
\begin{equation}\label{cond: beta_match1}
    \|\hat{\beta}^{AFS}_\rho(t)-\hat{\beta}^{LAR}(t)\|_2 \leq \epsilon,  
\end{equation}
which implies that the residuals $\hat{r}=y-X\hat{\beta}$ satisfy
\begin{equation}\label{cond: resid_match1}
    \|\hat{r}^{AFS}(t)-\hat{r}^{LAR}(t)\|_2 \leq \epsilon \|X\|_{op}.  
\end{equation}

Let $s^{(k)}(\rho)$ represent the AFS iteration when the $k$th unique variable enters the active set. To show the bound \ref{cond: beta_match1}, we show that for a sequence of $\epsilon_k >0$, there exists a $\rho \rightarrow 0$  such that 
\begin{equation}\label{cond: beta_match}
    \|\hat{\beta}^{AFS}_{s^{(k)}, \rho}-\hat{\beta}^{LAR}_{k}\|_2 \leq \epsilon_k \quad \forall k\in \{1,..,p\}
\end{equation} 
\ref{cond: beta_match} says that as $\rho \rightarrow 0$, the AFS coefficient is sufficiently close to that of LAR at the knots where a new variable enters the active set, since both LAR and AFS are piecewise linear functions.

Consider the base case at $k=0$. Both procedures initiate with $\hat{\beta}_0=0$ so $\epsilon_0=0$ and $\hat{r}^{AFS}_{s^{(0)}}=\hat{r}^{LAR}_{0}$. This also implies that $\mathcal{A}^{AFS}_{s^{(1)}}\equiv \mathcal{A}^{LAR}_{1}$. For the inductive step, assume that 
\begin{align*}
    \|\hat{\beta}^{AFS}_{s^{(k-1)}, \rho}-\hat{\beta}^{LAR}_{k-1}\|_2 \leq \epsilon_{k-1} \text{ and }
    \mathcal{A}^{AFS}_{s^{(k)}} \equiv \mathcal{A}^{LAR}_{k}.
\end{align*}
We now proceed with proving \ref{cond: beta_match} by induction. Let $d(\rho)\coloneqq s^{(k)}(\rho)-s^{(k-1)}(\rho)$ be the number of steps AFS takes before a new variable enters the active set. The coefficients in AFS updates as $\hat{\beta}^{\text{AFS}}_{m+1, \rho} = (1-\rho)\hat{\beta}^{\text{AFS}}_{m, \rho} + \rho \hat{\beta}^{OLS}_{\mathcal{A}_{m+1}}$ so expanding out the geometric sequence yields
\begin{equation}\label{afsupdate}
    \hat{\beta}^{\text{AFS}}_{s^{(k)}} = (1-\rho)^{d(\rho)}\hat{\beta}^{\text{AFS}}_{s^{(k-1)}} + \hat{\beta}^{OLS}_{\mathcal{A}_{s^{(k)}}}(1-(1-\rho)^{d(\rho)})
\end{equation}
Meanwhile, the update from LAR from step $k$ to $k+1$ can be written as
\begin{equation}\label{LARupdate}
    \hat{\beta}_{k+1}^{\text{LAR}} = (1-\zeta_{k+1})\hat{\beta}_{k}^{\text{LAR}} + \zeta_{k+1}\hat{\beta}^\text{OLS}_{\mathcal{A}_{k+1}}
\end{equation} 
where $\zeta_{k+1} \coloneqq \frac{\hat{\gamma}_{k+1}}{\bar{\gamma}_{k+1}} \in (0,1)$. Two properties of the procedures are key:

\begin{enumerate}
    \item  $\zeta_{k+1}$ is the smallest step in the direction of $(\hat{\beta}^\text{OLS}_{\mathcal{A}_{k+1}})_{j^*_{k+1}}$ before a new $j^* \notin \mathcal{A}_{k+1}$ enters the active set for the variable selection criterion. 
    \item In both LAR and AFS, once a new $j^*$ enters the active set at time $k+1$, no $j\in \mathcal{A}_{k}$ will be chosen again, by construction.
\end{enumerate} We will also use the following property, which follows from triangle inequality:

For any vectors $v_1$ and $v_2$ such that $\|v_1-v_2\| \leq c, c>0$ and any constants $a,b$, 
\begin{equation}\label{property}
    \|av_1-bv_2\|\leq a\|v_1-v_2\|+|a-b|\|v_2\|_2
\end{equation}
Let $\delta_k \coloneqq (1-(1-\rho)^{d(\rho)})-\zeta_k$ be the difference in the total distance AFS and LAR moves in the direction of a OLS coefficient before a new variable enters their respective active sets. Then we have
\begin{align}
    \|\hat{\beta}^{AFS}_{s^{(k)}, \rho}-\hat{\beta}^{LAR}_{k}\|_2 &\leq \|(1-\zeta_k)\hat{\beta}^{LAR}_{k-1}-(1-\rho)^{d(\rho)}\hat{\beta}^{\text{AFS}}_{s^{(k-1)}, \rho}\|_2 + \|(1-(1-\rho)^{d(\rho)}-\zeta_k)\hat{\beta}^{OLS}_{\mathcal{A}_{k}}\|_2\\
     &\leq |\delta_k|\|\hat{\beta}^{LAR}_{k-1}\|_2 + (1-\rho)^{d(\rho)}\|\hat{\beta}^{AFS}_{s^{(k-1)}}-\hat{\beta}^{LAR}_{k-1}\|_2+\delta_k\|\hat{\beta}^{OLS}_{\mathcal{A}_k}\|_2 \\
    &\leq 2|\delta_k|\|\hat{\beta}^{LAR}_{k-1}\|_2 + (1-\rho)^{d(\rho)}\epsilon_{k-1}\label{last_line}\
\end{align}
where the first inequality uses the representation of the AFS and LAR updates (\ref{afsupdate}, \ref{LARupdate}) and Cauchy-Schwarz while the second inequality follows by property \ref{property} and Cauchy-Schwarz. 

Now, we need to show that for any $k$, as $\rho\rightarrow 0$,
\begin{equation}\label{d_grow}
    d(\rho)\log(1-\rho) \rightarrow \log(1-\zeta_k)
\end{equation}
To show this, we have by the definition of $d(\rho)$ that
\begin{align*}
    &1-(1-\rho)^{d(\rho)-1} \leq \zeta_k \leq 1-(1-\rho)^{d(\rho)}
\end{align*}
and therefore 
\begin{align*}
    &\frac{d(\rho)-1}{d(\rho)} d(\rho)\log(1-\rho) \leq \log(1-\zeta_k) \leq d{(\rho)}\log(1-\rho)
\end{align*} 
Since $d(\rho)\rightarrow \infty$ as $\rho \rightarrow 0$, taking the limit as $\rho \rightarrow 0$ gives us \ref{d_grow}. 

Since $\|\hat{\beta}^{LAR}_{k-1}\|$ is fixed and not dependent on $\rho$, we have by \ref{d_grow} that as $\rho \rightarrow 0$, $\delta_k \rightarrow 0$. We also have $(1-\rho)^{d(\rho)} < 1$, giving us 
\begin{align*}
        \|\hat{\beta}^{AFS}_{s^{(k)}}-\hat{\beta}^{LAR}_{k}\|_2 &\leq  \epsilon_{k-1}
\end{align*}

when $\rho \rightarrow 0$. Since this holds for any arbitrary $k$, we can choose $\rho$ to be sufficiently small such that we can fix an $\epsilon_k$ that is arbitrarily close to $\epsilon_{k-1}$. By assumption, $\epsilon_{0}$ is 0, which proves \ref{cond: beta_match1}. 

So far, we have only shown that the LAR and AFS coefficients agree at the knots where a new variable enters the active set. To see that \ref{cond: beta_match} holds, recall that both procedures are piecewise linear functions moving in the direction of the OLS coefficient. Therefore, it must be that they also agree for any $t$ since those are points between the straight line that connects matching endpoints. 

Now, we prove that the active sets must also agree by showing that the difference between the selection criterion objective for AFS and LAR is arbitrarily small at any step.
By the assumption of no ties in variable selection, for any $\tilde{\epsilon}_k>0$,
\begin{equation}\label{resid_ineq}
    (x_{j^{*LAR}_k})^\intercal \hat{r}_{k-1}^{LAR} \geq \underset{j \in \{1,..p\}}{\text{ max }} x_j^\intercal \hat{r}_{k}^{LAR}+ \tilde{\epsilon}_k
\end{equation}
This inequality follows from the selection criterion of LAR,
$$j^{*LAR}_k = \underset{j \in \{1,..p\}}{\text{argmax }} x_j^\intercal \hat{r}^{LAR}_{k-1},$$
since the inner product of the selected variable column at step $k-1$ must be as large as the inner product of any variable column and the residual in the next step. $\epsilon_k$ can be arbitrarily close to $0$, so we can fix 
$$\epsilon_k \leq \frac{\tilde{\epsilon}_k}{\underset{j\in \{1,..p\}}{2\text{max }} \|x_j\|_2}.$$
Since AFS has the selection rule as LAR, the inequality \ref{resid_ineq} also holds for the AFS residual. This gives us
\begin{align*}
     |x_j^\intercal (\hat{r}_{s^{(k)}}^{AFS} - \hat{r}_{k}^{LAR})| & \leq |x_j^\intercal (\hat{r}_{s^{(k-1)}}^{AFS} - \hat{r}_{k-1}^{LAR})|\\
     &\leq |x_j^\intercal (X(\hat{\beta}_{s^{(k-1)}}^{AFS} - \hat{\beta}_{k-1}^{LAR}))|\\
     &\leq \epsilon_k \|X\|_{op}\\
     &\leq \tilde{\epsilon}_k,
\end{align*}
when combined with our inductive step assumption. Taking $\tilde{\epsilon}_k$ to $0$ concludes the proof.
\end{proof}
\subsection{Proof of Theorem \ref{theorem:thm2}}\label{app: thm2}

\begin{proof}

The below derivation draws from the proof of Theorem 2 in \cite{sparseboost}. We can equivalently write \ref{eqn: ortho_betaj} as
$$\hat{\beta}^{AFS}_m = D_m X^\intercal  y = D_m \hat{\beta}^{OLS}$$
where $D_m \in \mathbbm{R}^{p\times p}$ is a diagonal matrix with the $j$-th entry as $(1-(1-\rho)^{\ell_{j,m}})$.
Then $\Delta_{m-1, m} \coloneqq RSS_{m}-RSS_{m-1}$ decreases in $m$ such that $\Delta_{m-1, m} = (1-\rho)^2\Delta_{m, m+1}$ since
\begin{align*}
	\Delta_{m-1, m} &= \|y-XD_mX^\intercal y\|^2_2-\|y-XD_{m-1}X^\intercal y\|^2_2 \\
	&=\|X^\intercal (y-XD_mX^\intercal y)\|^2_2-\|X^\intercal (y-XD_{m-1}X^\intercal y)\|^2_2\\
	&=\|(I-D_m)X^\intercal y\|^2_2-\|(I-D_{m-1})X^\intercal y\|^2_2\\
 &= \sum_{j=1}^p ([(I-D_m)^2-(I-D_{m-1})^2](\hat{\beta}^{OLS})^2_j\|_2.
\end{align*}
Here, we apply the orthogonality of $X$ to get the third equality.
Then for each fixed $j\in \mathcal{A}$, $\exists \delta^2 >0$ such that 
\begin{align*}
    &\left[(1-\rho)^{2(\ell_{j,h})}-(1-\rho)^{2(\ell_{j,h+1})} \right](\hat{\beta}^{OLS})^2_j > \delta^2_{h,j}, \ \ h \in \{1,..,m-1\} \\
    &\left[(1-\rho)^{2(\ell_{j,m+1})}-(1-\rho)^{2(\ell_{j,m+1})} \right](\hat{\beta}^{OLS})^2_j 
    \leq \delta^2_{m,j}
\end{align*}

From the above, consider the approximation $\delta^2_{m,j} \approx (1-\rho)^{2\ell_{j,m}}(1-(1-\rho)^2)(\hat{\beta}^{OLS})^2_j$. This gives us the following soft-threshold approximation:

\begin{equation}
     \hat{\beta}^{AFS}_{j,m} \approx \hat{\beta}^{ST}_{j,m} \coloneqq
    \begin{cases}
      \hat{\beta}^{OLS}_j - \lambda_{j,m} & \text{if } \hat{\beta}^{OLS}_j \geq \lambda_{j,m}\\
      0 & \text{if } |\hat{\beta}^{OLS}_j| < \lambda_{j,m}\\
      \hat{\beta}^{OLS}_j + \lambda_{j,m} & \text{if } \hat{\beta}^{OLS}_j \leq -\lambda_{j,m}
    \end{cases}       
\end{equation}

where $\lambda_{j,m} = \frac{\delta_{m,j}}{\sqrt{1-(1-\rho)^2)}}$. In fact, when $\rho \rightarrow 0$, the difference in the estimators goes to $0$. We refer readers to \cite{sparseboost} for details. Finally, since the largest possible $\Delta_{m-1,m}$ occurs when $m=k_j$, we get that the value of $\lambda_{j,m}$ must be between $$(1-\rho)^{2\ell_{j,m}}\sqrt{1-(1-\rho)^2} \text{ and } (1-\rho)^{2}\sqrt{1-(1-\rho)^2}$$
\end{proof}

\begin{figure}[H]
    \centering
    \captionsetup{width=\linewidth}
    \includegraphics[width=\linewidth]{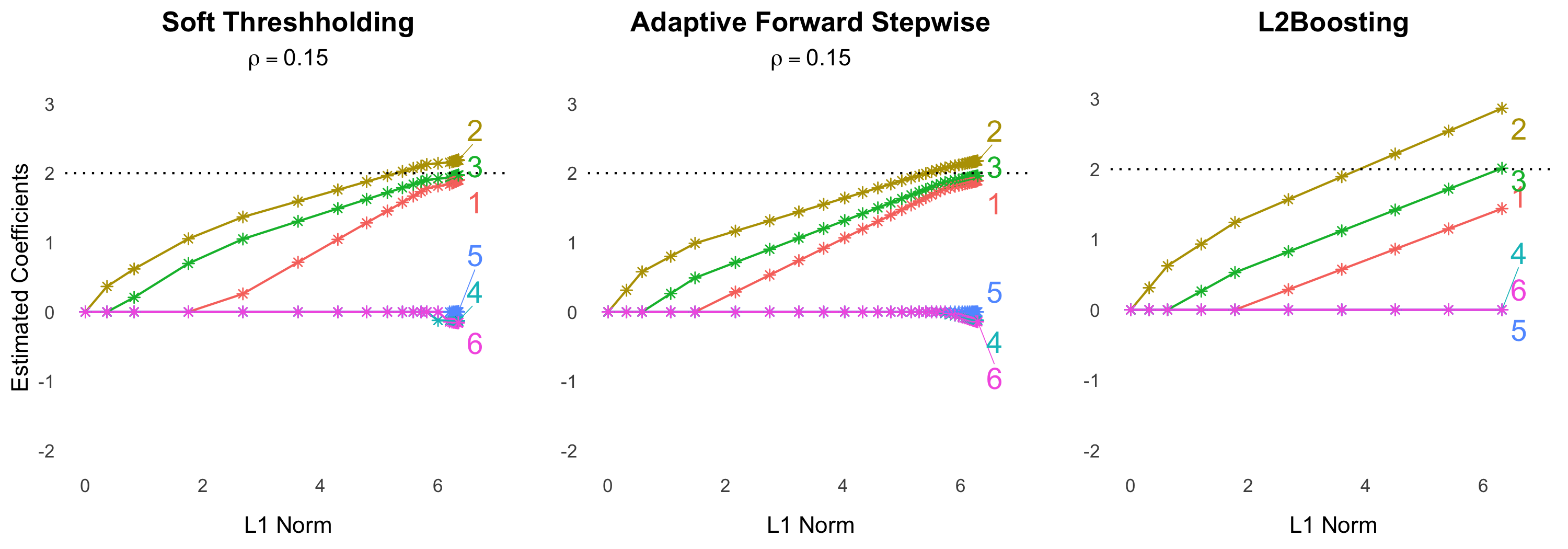}
    \caption{Figure: \ref{fig: non-ortho}: Coefficient paths for the approximate soft thresholding estimator \ref{eqn: soft} at $\rho=0.15$, AFS at $\rho=0.15$, and L2Boosting. Simulation for $n=100, p=6$ under a non-orthogonal design matrix $X$, with $\beta_1, \beta_2, \beta_3 = 2$ and $0$ otherwise.}
    \label{fig: non-ortho}
\end{figure}

\begin{remark}
    Although the monotonic behavior of $\Delta_{m-1,m}$ no longer holds under non-orthogonal design, $RSS_m$ is still decreasing in $m$. As a result, for small $\rho$, we would expect \ref{eqn: soft} to still be a reasonable estimator, as seen empirically in Figure \ref{fig: non-ortho} if $\Delta_{m-1,m}$ is decreasing in $m$ for most iterations. 
\end{remark}

\section{Simulations}

\subsection{Degrees of freedom}\label{app:dof}

We use $B=1000$ bootstrap iterations to estimate the dof $\sum_{i=1}^n \text{Cov}(y_i, \hat{y}_i)/\sigma^2$ \citep{dof}. As anticipated, the dof varies between that of the LASSO ($\text{dof} = |\mathcal{A}|$) and FS, as seen in \cite{bestsubset}.
\begin{figure}[H]
    \centering
    \captionsetup{width=\linewidth}
    \includegraphics[width=0.5\linewidth]{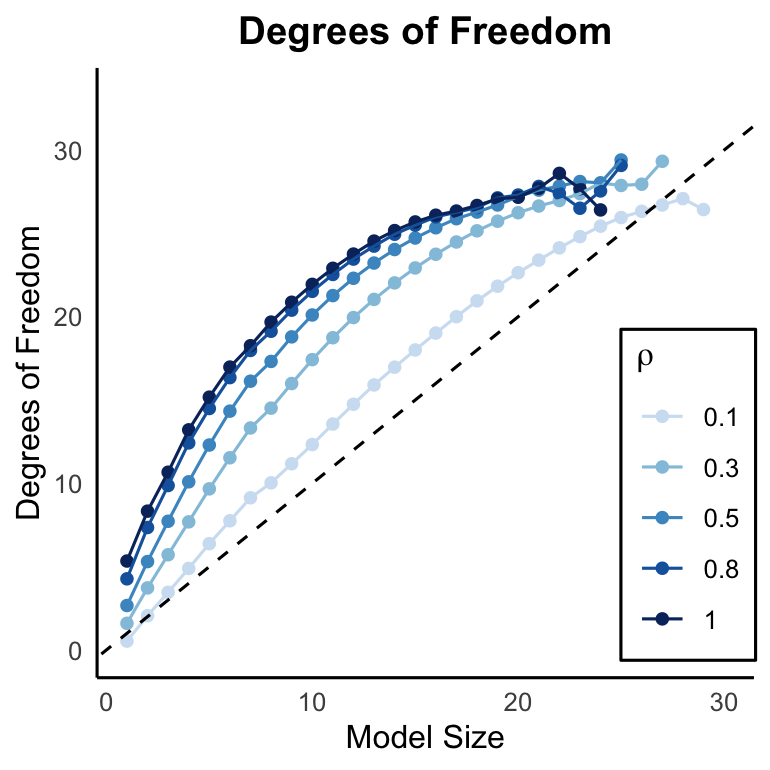}
    \caption{Bootstrap estimates of dof for AFS under different $\rho$. Simulation uses $\sigma=1.75$}
    \label{fig:enter-label}
\end{figure}
\label{appendix:placeholder}

\subsection{Benchmark results}

\begin{figure}[H]
    \centering
    \captionsetup{width=\linewidth}
    \includegraphics[width=\linewidth]{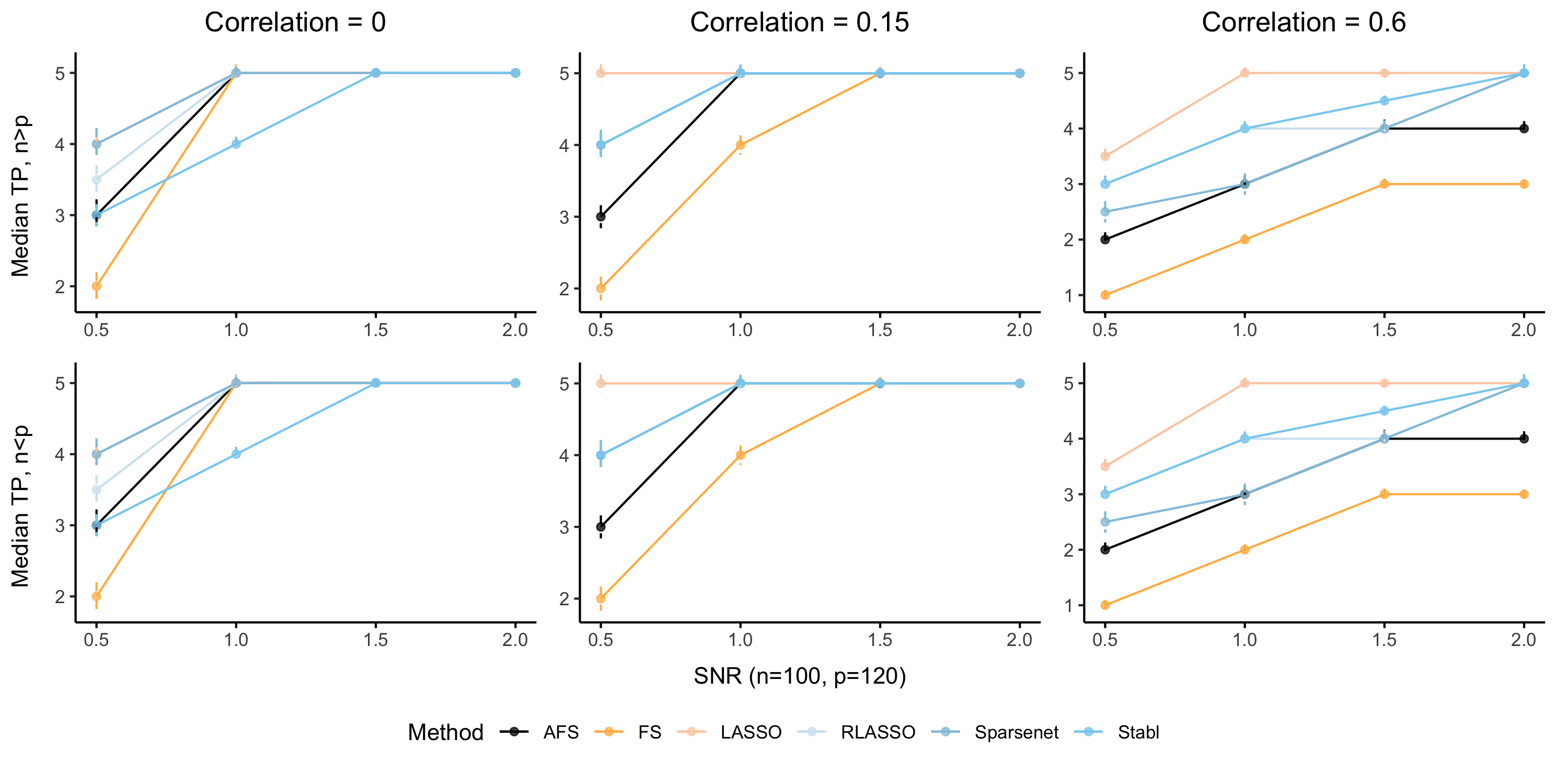}
    \caption{Figure \ref{app: TP}: Simulation results averaged across $50$ trials. AFS recovers the true support under no and low correlation settings. However, it tends to perform less well in the lowest SNR and high correlation setting.}
    \label{app: TP}
\end{figure}

\section{Inference}\label{app:inf}

We are interested in the following test
\begin{align*}
    H_0: v^\intercal \theta = 0
\end{align*}
conditional on the chosen $A_k$ at step k. Then the selection event, $y$ of AFS at step $k$ can be represented as a polyhedron of the form 
$$ \mathcal{P} = \{y: \hat{A}_k = A_k, \hat{s}_{A_k}=s_{A_k}\} = \{y: \Gamma y > 0\}$$ where $A_k = [j_1, ..., j_k]$ is the set of active variables after $k$ steps, $s_k=\text{sign}(X_k^\intercal r)$, and $r$ is the residual from regressing $y$ onto $X_{A_{k-1}}$. The construction follows that of the polyhedral sets for FS selection events as in \cite{ryan_postselection}. By induction, we get the matrix $\Gamma$ with rows based on the following conditions:
\begin{align*}
    s_kX_{j_k}^\intercal \bigg(I-X\rho \sum_{i=0}^{k-1} (1-\rho)^k \hat{\nu}_{A_{k-1-i}}\bigg)y &\geq \pm X_j^\intercal \bigg(I-X\rho \sum_{i=0}^{k-1} (1-\rho)^k \hat{\nu}_{A_{k-1-i}}\bigg)y, \ \forall j \neq j_k \\
     s_kX_{j_k}^\intercal \bigg(I-X\rho \sum_{i=0}^{k-1} (1-\rho)^k \hat{\nu}_{A_{k-1-i}}\bigg)y &> \pm X_j^\intercal \bigg(I-X\rho \sum_{i=0}^{k-1} (1-\rho)^k \hat{\nu}_{A_{k-1-i}}\bigg)y, \ \forall j \in A_{k-1} 
\end{align*}

Unlike in FS, the additional condition is needed to guarantee we are looking at the set for which $X_{j_k}$ gets chosen for the first time. Then we can directly apply Theorem 5.2 of \cite{lee_postselection} to get a conditional test statistic. We refer readers to the paper for details on the test and confidence interval construction.

\section{Modification of AFS to recover LAR and the LASSO}\label{lasso_mod}

We can modify the AFS algorithm to recover LAR for a fixed $\rho$:

\begin{algorithm}[H]    
\caption{Adaptive Forward Stepwise - LAR Modifications}\label{alg:afs_lars}
\begin{algorithmic}[1]
    \addtocounter{ALG@line}{5}
     \State While $m < M$ and $\|\hat{\beta}^{AFS}\|_1 < h$, let
        \State \quad  $ j^*_m= \underset{j \in \{1,..,p\}}{\text{argmax }} |x_j^\intercal ({y}- X\hat{\beta}^{AFS}_{m-1} )|$ \codecomment{Select most correlated variable with current residuals} 
        \State \quad $A_m=A_{m-1} \cup j^*_m$ \codecomment{Update active set}
        \State \quad  $\hat{\nu}_m=\hat{\beta}^{OLS}_{\mathcal{A}_m}$, the OLS coefficients of $y$ on the active set $\mathcal{A}_m$ \codecomment{Compute OLS coefficients}
        \State \quad Begin recovery of LAR coefficient:
        \State \quad \quad Fix $\epsilon_m > 0$ small and let $\tilde{\rho} = \rho - \epsilon_m$
            \State \quad \quad $\tilde{\beta}^{AFS} = (1 - \tilde{\rho}) \hat{\beta}^{AFS}_{m-1} + \tilde{\rho} \hat{\nu}_m$
            \State \quad \quad$\tilde{j} = \underset{j \in \{1,...,p\}}{\text{argmax }} |x_j^\intercal ({\bf{y}} - \sum_{p=1}^{P} \tilde{\beta}^{AFS}_{k-2, p} x_p)|$
            \State \quad \quad While $\tilde{j}= j_m^*$
                \State  \quad \quad \quad Increase $\epsilon_m$ and repeat LAR coefficient recovery. As $\tilde{\rho} \rightarrow \rho$, recovers the LAR coefficient
        \State \quad $\hat{\beta}^{AFS}_m=(1-\tilde{\rho}) \hat{\beta}^{AFS}_{m-1} + \tilde{\rho} \hat{\nu}_m$  \codecomment{Update AFS coefficients} 
\end{algorithmic}
\end{algorithm}
We can further modify the above to recover the LASSO with the same restriction as in \cite{lars}: If $\text{sign}(\hat{\beta}^{AFS}_{k,j}) \neq \text{sign}(\hat{\beta}^{AFS}_{k-1,j}) \neq 0$, remove $x_j$ from $\mathcal{A}_k$ and repeat procedure.

\end{document}